\documentclass[runningheads]{llncs}
\usepackage[T1]{fontenc}
\usepackage{amsmath}
\usepackage{multicol}
\usepackage{amsfonts}
\usepackage{xcolor}
\usepackage{graphicx}
\usepackage{tcolorbox}
\usepackage[ruled,linesnumbered]{algorithm2e}

\usepackage{url}
\usepackage{hyperref}
\hypersetup{
    colorlinks=true,
    linkcolor=blue,
    filecolor=magenta,      
    urlcolor=cyan,
}
\begin{document}
\title{Influential Billboard Slot Selection under Zonal Influence Constraint}
\author{Dildar Ali \and Suman Banerjee \and Yamuna Prasad}
\authorrunning{Ali et al.} 
\institute{Department of Computer Science and Engineering,\\
Indian Institute of Technology Jammu \\
Jammu \& Kashmir-181221, India.\\
\email{2021rcs2009@iitjammu.ac.in, suman.banerjee@iitjammu.ac.in, yamuna.prasad@iitjammu.ac.in}}
\maketitle
\begin{abstract}
Given \emph{billboard} and \emph{trajectory database}, finding a limited number of billboard slots for maximizing the influence is an important problem in the context of \emph{billboard advertisement}. Most of the existing literature focused on the influential slot selection problem without considering any specific zonal influence constraint. To bridge this gap in this paper, we introduce and study the Influential Billboard Slot Selection Problem Under Zonal Influence Constraint. We propose a simple greedy approach to solve this problem. Though this method is easy to understand and simple to implement due to the excessive number of marginal gain computations, this method is not scalable. We design a branch and bound framework with two bound estimation techniques that divide the problem into different zones and integrate the zone-specific solutions to obtain a solution for the whole. We implement both the solution methodologies with real-world billboard and trajectory datasets and several experiments have been reported. We compare the performance of the proposed solution approaches with several baseline methods. The results show that the proposed approaches lead to more effective solutions with reasonable computational overhead than the baseline methods. 
\end{abstract}

\section{Introduction}
\emph{Billboard Advertising} has become effective out-of-home marketing strategy, offering excellent returns on investment \footnote{\url{https://topmediadvertising.co.uk/billboard-advertising-statistics/}}. Several commercial houses have emerged as effective players in this domain. This area has grown tremendously in the last decade, and by the end of 2024, it has become $410.82$ billion dollar industry \footnote{\url{https://www.thebusinessresearchcompany.com/report/billboard-and-outdoor-advertising-global-market-report}}. To make an effective billboard advertisement, it is always important to choose most billboard slots that will reach a large number of people. However, each billboard slot is associated with some cost, which means if a commercial house hires a particular slot, a pre-specified amount of money needs to be paid to the influence provider. As per the marketing literature reports, a commercial house spends around $7-10 \%$ of its annual revenue for advertising purposes\footnote{\url{https://www.lamar.com/howtoadvertise/Research/}}. So, it is an important question: given the information about billboard slots and corresponding costs, how can we select a limited number of them? This problem is called the \textsc{Billboard Slot Selection} Problem.

\paragraph{\textbf{Background.}} To the best of our knowledge, Zhang et al. \cite{10.1145/3219819.3219946} were the first to study the billboard advertisement problem. Further, Zhang et al. \cite{10.1145/3292500.3330829} find a set of billboards with maximum influence under a budget when impression counts are considered. Zhang et al. \cite{zhang2020towards} further extend their previous work \cite{10.1145/3219819.3219946} to find a set of most influential billboards under a budget constraint. Wang et al.\cite{8604082} find $k$ best advertising units which maximize the influence spread in targeted advertising. Similarly, Wang et al.\cite{wang2022data} study the problem of targeted influence maximization under a limited budget constraint. Deviating from the influence maximization, some literature exists \cite{ali2024minimizing,Ali_Bhagat_Banerjee_Prasad_2023,10.1145/3448016.3457257} on regret minimization in the context of billboard advertisements. To the best of our knowledge, none of the studies have considered the budget associated with a billboard slot, unlike the study by Ali et al.\cite{10.1007/978-3-031-22064-7_17,ali2023influential}. Also, none of the studies have considered the concept of zonal influence constraint. Now, we describe the motivation behind this constraint. 

\paragraph{\textbf{Motivation.}} In all the previous studies \cite{10.1145/3219819.3219946,10.1007/978-3-031-22064-7_17,10.1145/3292500.3330829,zhang2020towards}, 
It is assumed that each zone in a city is equally meaningful to an advertiser during the selection of slots. However, this is not always true in real-life scenarios \cite{ali2024minimizing}. Let's consider the zone-specific population information of a city is known to the advertiser. Now, it is meaningless to advertise a highly expensive product in a demographic zone where economically backward classes people live. Because they may not buy the product, investing money to advertise such a product by the advertiser is totally waste. Hence, we divide the city into several zones based on the demographic information. Now, the advertisers have zonal influence demand along with the budget, and the influence provider must fulfill this. So, studying the \textsc{Billboard Slot Selection Problem} under zonal influence constraint is important.

\paragraph{\textbf{Our Contributions.}} This paper introduces the problem of selecting influential billboard slots under zonal influence constraint. We call this problem \textsc{Influential Billboard Slot Selection under Zonal Influence Constraint (IBSSP)}. To the best of our knowledge, this is the first study in this direction. The key contributions of this paper are as follows:
\begin{itemize}
    \item We introduce and formulate this problem as a discrete optimization problem and show this problem as NP-hard to approximate within a constant factor.
    \item We propose a simple greedy and branch and bound approach with two bound estimation techniques to solve this problem.
    \item  We substantiate our claim regarding the efficiency and effectiveness of the proposed solutions by experimental evaluations with real-world datasets.
\end{itemize}
\paragraph{\textbf{Organization of the Paper.}} The rest of the paper is organized as follows. Section \ref{Sec:BPD} describes the background of our problem and defines it formally. The proposed solution approaches have been described in Section \ref{Sec:PSA}. Section \ref{Sec:EE} contains the experimental evaluation of the proposed solution approaches. Finally, Section \ref{Sec:Conclusion} concludes our study and mentions future research directions.

\section{Background and Problem Definition} \label{Sec:BPD}
Consider a set of $m$ digital billboards $\mathbb{B}=\{b_1, b_2, \ldots, b_m\}$ are placed across a geographical region of interest. Each one is operated for the duration $[T_1, T_2]$ and we denote $T=T_2-T_1$. Each billboard is allocated slot-wise with the duration of the slot equal to $\Delta$. Now, we define the notion of a billboard slot.
\begin{definition} [Billboard Slot]
    Given a set of billboards and a fixed duration of slot $\Delta$, a billboard slot is defined by a tuple $(bs_i, [t,t+\Delta])$ where $bs_i \in \mathbb{B}$ and $t \in \{T_1, T_1 + \Delta+1, T_1 + 2\Delta+1, \ldots, T_2 - \Delta+1\}$.
\end{definition}
It can be observed that the number of billboard slots will be equal to $m \cdot \frac{T}{\Delta}$. Now, each billboard slot is associated with some cost, and this is formalized by the cost function $\mathcal{C}$ that maps each billboard slot to its associated cost $\mathcal{C}(bs_{i}) ~\forall~ bs_{i} \in \mathcal{BS}$, i.e., $\mathcal{C}: \mathcal{BS} \longrightarrow \mathbb{Z}^{+}$. Now, a subset of the slots can be chosen with a fixed budget of $\mathcal{B}$. Next, we define the influence of a subset of billboard slots in Definition \ref{Def:Influence}.
\begin{definition} [Influence of Billboard Slots] \label{Def:Influence}
    Given a trajectory and billboard database and a subset of billboard slots $\mathcal{S} \subseteq \mathcal{BS}$, the influence of the billboard slots is given by Equation No. \ref{Eq:Influence}.

    \begin{equation} \label{Eq:Influence}
        \mathcal{I}(\mathcal{S})=\underset{u_j \in \mathcal{U}}{\sum} \ [1- \underset{bs_{i} \in \mathcal{S}}{\prod} (1-Pr(bs_{i},u_j))]
    \end{equation}
\end{definition}
Here, $\mathcal{I}()$ is the influence function that maps all possible subsets of the billboard slots to its expected influence, i.e., $\mathcal{I}:2^{\mathcal{BS}} \longrightarrow \mathbb{R}_{0}^{+}$ with $\mathcal{I}(\emptyset)=0$. Now, consider the geographical region under consideration is denoted by $\mathcal{Z}$, and this is divided into $\ell$ many zones $\mathcal{Z}=\{z_1, z_2, \ldots, z_{\ell}\}$. Consider an advertiser $a_{i}$ wants to choose a subset of billboard slots within the allocated budget $\mathcal{B}$ such that zonal influence demand is satisfied. Let, for any zone $z_i \in \mathcal{Z}$ the influence requirement is $\sigma_{i}$. The influence of the billboard slot subset $\mathcal{S}$ subject to the zone $z_i$ is denoted by $\mathcal{I}_{z_i}(\mathcal{S})$ and $\mathcal{I}(\mathcal{S})=\underset{z_i \in \mathcal{Z}}{\sum} \ \mathcal{I}_{z_i}(\mathcal{S})$. So the problem becomes: `` Given a billboard database $\mathbb{B}$, trajectory database $\mathbb{T}$ and a cost function $\mathcal{C}$, the problem of Influential Billboard Slot Selection with Zonal Influence Constraint asks to choose a subset of the billboard slots $\mathcal{S}$ such that the total selection cost is less than the budget, i.e., $\underset{b_i \in \mathcal{S}}{\sum} \mathcal{C}(b_i) \leq \mathcal{B}$ and the zonal influence constraint for all the zones are satisfied, i.e., $\mathcal{I}_{z_i}(\mathcal{S}) \geq \sigma_{i}$ for all $i \in [\ell]$ such that the influence is maximized". Now, we define the zonal influence constraint in Definition \ref{ZIC}.

\begin{definition} [Zonal Influence Constraint]\label{ZIC}
Let $\mathbb{BS}_{a_i}$ be the set of allocated billboard slots to the advertiser $a_i$ and $\mathbb{BS}^{z_j}_{a_i}$ $\subseteq$ $\mathbb{BS}_{a_i}$, contains the billboard slots associated with zone $z_j$. Now, we can say this allocation satisfies the zonal influence constraint for zone $z_j$ if and only if $\mathcal{I}(\mathbb{BS}^{z_j}_{a_i}) \geq \sigma^{j}_{i}$.  
\end{definition}
Next, this problem can be formally stated as follows:
\begin{center}
\begin{tcolorbox}[title=\textsc{Slot Selection in Billboard Advertisement} Problem, width=11.6cm]
\textbf{Input:} Billboard Slots set $\mathcal{BS}$, Trajectory Database $\mathbb{T}$, Influence Function $\mathcal{I}()$, Budget $\mathcal{B}$, Advertiser Database $\mathcal{A}$.

\textbf{Problem:} Find out a subset of billboard slots $\mathcal{P} \subseteq \mathcal{BS}$, such that zonal influence demand is satisfied and total influence is maximized.
\end{tcolorbox}
\end{center}

The influence maximization problem without considering zonal influence constraints in billboard advertisement has been studied by Zhang et al. \cite{zhang2020towards}, and they had inapproximability results stated in Theorem \ref{Theorem1}.
\begin{theorem} \label{Theorem1}
    \textsc{The Influential Billboard Slot Selection Problem Under Zonal Influence Constraint} is NP-hard and hard to approximate with a constant factor algorithm.
\end{theorem}
 This means the same inapproximability result also holds for our problem. Next, in section \ref{Sec:PSA}, we discuss the proposed solution methodologies.

\section{Proposed Solution Approach} \label{Sec:PSA}

\subsection{Simple Greedy Allocation Scheme.} This kind of greedy approach has already been studied by Zhang et al. \cite{zhang2020towards} for influential billboard selection problems. Hence, we have extended the greedy approach for our problem context, i.e., under zonal influence constraint. A simple straight forward greedy approach for selecting billboard slots is defined in Algorithm \ref{Algo:Greedy} (Line No. $1$ to $22$), which selects the billboard slots that maximize the marginal influence gain to a candidate solution set $\mathcal{P}$ to fulfill the zone-specific influence demand of an advertiser until the budget $(\mathcal{B})$ is exhausted. However, a simple greedy approach does not guarantee the approximation ratio due to budget constraints. Hence, Line No. $23$ to $49$ is added with simple greedy to overcome the drawback and guarantees an approximation ratio. Basically, the naive greedy approach is used where the billboard slots are chosen based on the individual's highest influence value to solve the \textsc{IBSSP} problem.
\begin{algorithm}[ht!]
\scriptsize
\SetAlgoLined
\KwData{Trajectory Database $\mathbb{T}$, Billboard Database $\mathbb{B}$, Advertiser Database $\mathcal{A}$, and the Influence Function $\mathcal{I}()$, Zone Information, Budget $\mathcal{B}$, Zonal influence demand $\sigma$.}
\KwResult{A Billboard slot set $\mathcal{P} \subseteq \mathbb{BS}$ such that $\mathcal{C}(\mathcal{P}) \leq \mathcal{B}$}
\begin{multicols}{2}
$\text{Initialize } \mathcal{K} \leftarrow \emptyset$, $\mathcal{P} \leftarrow \emptyset$, $\mathcal{Q} \leftarrow \emptyset$, $\mathcal{B}_{1} = \mathcal{B}_{2} = \mathcal{B}$\;

\For {each $z_{j} \in a_{z}$}{
    $\mathcal{S} \leftarrow \emptyset$\;

    \While{$\sigma_{z_j} > I(\mathcal{S}) \text{ and } \mathcal{BS}_{z_j} \neq \emptyset$}{
        $s^{*} \leftarrow \underset{s \in \mathcal{BS}_{z_j}}{\text{argmax}} \ \frac{\mathcal{I}(\mathcal{S} \cup \{s\}) - \mathcal{I}(\mathcal{S})}{\mathcal{C}(\{s\})}$\;

        \If {$\mathcal{C}(\mathcal{S}) +  \mathcal{C}(\{s^{*}\}) \leq \mathcal{B}_{1}$}{
            $\mathcal{S} \leftarrow \mathcal{S} \cup \{s^{*}\}$\;
            $\mathcal{BS}_{z_j} \leftarrow \mathcal{BS}_{z_j} \setminus \{s^{*}\}$\;
            $\mathcal{B}_{1} \leftarrow \mathcal{B}_{1} - \mathcal{C}(s^{*})$\; 
        }
    }
    $\mathcal{P} \leftarrow \mathcal{P} \cup \{\mathcal{S}\}$\;   
}
$\mathcal{BS} \leftarrow \{\mathcal{BS}_{z_1} \cup \mathcal{BS}_{z_2} \cup \ldots \cup \mathcal{BS}_{z_l}\}$\;
\While{$\mathcal{B}_{1} > 0 \text{ and } \mathcal{BS} \neq \emptyset$}{
    $s^{*} \leftarrow \underset{s \in \mathcal{BS}}{\text{argmax}} \ \frac{\mathcal{I}(\mathcal{S} \cup \{s\}) - \mathcal{I}(\mathcal{S})}{\mathcal{C}(\{s\})}$\;

    \If {$\mathcal{C}(\mathcal{P}) +  \mathcal{C}(\{s^{*}\}) \leq \mathcal{B}_{1}$}{
        $\mathcal{P} \leftarrow \mathcal{P} \cup \{s^{*}\}$\;
        $\mathcal{BS} \leftarrow \mathcal{BS} \setminus \{s^{*}\}$\;
        $\mathcal{B}_{1} \leftarrow \mathcal{B}_{1} - \mathcal{C}(s^{*})$\;
    }
}

\columnbreak

\For {each $z_{j} \in a_{z}$}{
    $\mathcal{K} \leftarrow \emptyset$\;

    \While{$\sigma_{z_j} > I(\mathcal{K}) \text{ and } \mathcal{BS}_{z_j} \neq \emptyset$}{
        $k^{*} \leftarrow \underset{s \in \mathcal{BS}_{z_j}}{\text{argmax}} \ \mathcal{I}(\mathcal{K} \cup \{s\})$\;

        \If {$\mathcal{C}(\mathcal{K}) +  \mathcal{C}(\{k^{*}\}) \leq \mathcal{B}_{2}$}{
            $\mathcal{K} \leftarrow \mathcal{K} \cup \{k^{*}\}$\;
            $\mathcal{BS}_{z_j} \leftarrow \mathcal{BS}_{z_j} \setminus \{k^{*}\}$\;
            $\mathcal{B}_{2} \leftarrow \mathcal{B}_{2} - \mathcal{C}(k^{*})$\; 
        }
    }
    $\mathcal{Q} \leftarrow \mathcal{Q} \cup \{\mathcal{K}\}$\;  
  
}
$\mathcal{BS} \leftarrow \{\mathcal{BS}_{z_1} \cup \mathcal{BS}_{z_2} \cup \ldots \cup \mathcal{BS}_{z_l}\}$\;
  \While{$\mathcal{B}_{2} > 0 \text{ and } \mathcal{BS} \neq \emptyset$}{
        $k^{*} \leftarrow \underset{s \in \mathcal{BS}}{\text{argmax}} \ \mathcal{I}(\mathcal{K} \cup \{s\})$\;

        \If {$\mathcal{C}(\mathcal{Q}) +  \mathcal{C}(\{k^{*}\}) \leq \mathcal{B}_{2}$}{
            $\mathcal{Q} \leftarrow \mathcal{Q} \cup \{k^{*}\}$\;
            $\mathcal{BS} \leftarrow \mathcal{BS} \setminus \{k^{*}\}$\;
            $\mathcal{B}_{2} \leftarrow \mathcal{B}_{2} - \mathcal{C}(k^{*})$\;
        }
    }
\If {$\mathcal{I}(\mathcal{Q}) > \mathcal{I}(\mathcal{P})$}{
    return $\mathcal{Q}$\;
}
\Else{
    return $\mathcal{P}$
}
\end{multicols}
\caption{Simple Greedy Approach for Influential Billboard Slot Selection Problem}
\label{Algo:Greedy}
\end{algorithm}
\paragraph{\textbf{Complexity Analysis.}}
In-Line No. $1$ initializing an empty sets $\mathcal{K}$,$\mathcal{P}$,$\mathcal{Q}$ will take $\mathcal{O}(1)$ time. In-Line No. $2$ \texttt{for loop} and Line No. $4$ \texttt{while loop} will execute for $\mathcal{O}(\ell)$ and $\mathcal{O}(m)$ time, respectively. In-Line No. $5$, computing influence will take $\mathcal{O}(2 \cdot \ell \cdot m \cdot t)$ where $t$ is the number of tuples in the trajectory database. Line No. $6$ to $10$ will execute for $\mathcal{O}(\ell \cdot m)$  time. Hence, Line No. $1$ to $14$ will take $\mathcal{O}(m^{2} \cdot \ell \cdot t^{2})$. Similarly, Line No. $15$ to $22$ will take $\mathcal{O}(m^{2} \cdot t)$. In Line No. $23$ to $35$ will take $\mathcal{O}(m^{2} \cdot \ell \cdot t^{2})$ and Line No. $36$ to $49$ will take $\mathcal{O}(m^{2} \cdot t)$ time to execute. So, Algorithm \ref{Algo:Greedy} will take total $\mathcal{O}(m^{2} \cdot \ell \cdot t^{2} + m^{2} \cdot t)$ i.e., $\mathcal{O}(m^{2} \cdot \ell \cdot t^{2})$ time to execute. The extra space taken by Algorithm \ref{Algo:Greedy} will be of $\mathcal{O}(\ell + m)$. 

\subsection{Branch and Bound Approach.}
 Although Simple Greedy achieves a constant approximation guarantee, executing requires huge amounts of computational time. To overcome the drawback, we introduce the Branch and Bound approach in Algorithm \ref{Algo:BB}. We first initialize $U_{G}$, $L_{G}$ and the max heap $(\mathcal{P}, \Bar{\mathcal{P}}, \mathcal{Y},\mathcal{U})$, where $\mathcal{P}$ is the solution set, $\Bar{\mathcal{P}}$ is the set of unexplored billboard slots, $\mathcal{Y}$ is the zone-specific influence demand, and $\mathcal{U}$ is the upper bound of the influence. The max heap $\mathcal{H}$ is ordered based on the upper bound of influence $\mathcal{U}$. Hence, the heap will pop the top entry, which has a maximum $\mathcal{U}$ value till $L_{G} < U_{G}$. Further, as long as the budget constraint is satisfied, it will generate two new branches $\mathcal{P}^{a}$ and $\mathcal{P}^{b}$. Initially, $\mathcal{P}^{a}$ denotes the feasible slots set by adding $b$ while $\mathcal{P}^{b}$ is the feasible slots set excluding $b$ (Lines $9$ - $11$). Now, based on the value of $\mathcal{P}^{a}$ (or $\mathcal{P}^{b}$) and $\Bar{\mathcal{P}}$ value the $BoundEstimation(.)$ or $FastBoundEstimation$ function will return $(\mathcal{P}^{c}, L^{a}, \mathcal{Y},\mathcal{U}^{a})$ or $(\mathcal{P}^{c}, L^{b},\mathcal{Y},\mathcal{U}^{b})$. In-Line No. $13$ if the current satisfied influence demand $(L^{a})$ is greater than previous satisfied demand $(L^{G})$ then update $L^{G}$ with $L^{a}$ and $\Hat{\mathcal{P}}$ with $\mathcal{P}^{c}$ as it may be the possible solution set. Next, in Line No. $16$, if updated influence demand $(\mathcal{U}^{a})$ still greater than $L_{G}$, then insert that node into the max heap $\mathcal{H}$ for further explorations. We do the same in Line No. $19$ to $25$ for $\mathcal{S}^{b}$. We repeat the $\texttt{while loop}$ for all the branches until $L_{G} \geq U_{G}$.

\SetKwComment{Comment}{/* }{ */}
\begin{algorithm}[h]
\scriptsize
\SetAlgoLined
\KwData{Trajectory Database $\mathbb{T}$, Billboard Database $\mathbb{B}$, and the Influence Function $I()$, Zonal Demand $\mathcal{Y}$, Zone Info. $a_{z}$, Budget $\mathcal{B}$.}
\KwResult{A Billboard slot set $\Hat{\mathcal{P}} \subseteq \mathbb{BS}$ such that $\mathcal{C}(\Hat{\mathcal{P}}) \leq \mathcal{B}$}

Initialize $\Hat{\mathcal{P}} \leftarrow \emptyset$, $\mathcal{P} \leftarrow \emptyset$, $\Bar{\mathcal{P}} \leftarrow \mathbb{BS},\mathcal{Y}= (\sigma_{z_{1}},\sigma_{z_{2}}, \ldots \sigma_{z_{\ell}})$\;
Initialize max heap $\mathcal{H} \leftarrow (\mathcal{P}, \Bar{\mathcal{P}}, \mathcal{Y},\mathcal{U})$\;

$L_{G} \leftarrow 0$, $U_{G} \leftarrow \infty$\;
\While{$L_{G} < \theta . U_{G}$}{
$(\mathcal{P}, \Bar{\mathcal{P}}, \mathcal{Y},\mathcal{U}) \leftarrow$ top of $\mathcal{H}$\;
$U_{G} \leftarrow \mathcal{U}$\;
\For{each $b \in \Bar{\mathcal{P}}$}{
\If{$\mathcal{C}(\mathcal{P})$ + $\mathcal{C}(b) \leq \mathcal{B}$}{
$\Bar{\mathcal{P}} \leftarrow \Bar{\mathcal{P}} \setminus \{b\}$\;
$\mathcal{P}^{a} \leftarrow \mathcal{P} \cup \{ b\}$\;
$\mathcal{P}^{b} \leftarrow \mathcal{P}$\;
$(\mathcal{P}^{c}, L^{a}, \mathcal{Y},\mathcal{U}^{a}) \leftarrow \text{BoundEstimation}(\mathcal{P}^{a}, \Bar{\mathcal{P}}, a_{z}, \mathcal{Y})$\;
\If{$L^{a} > L_{G}$}{
$L_{G} \leftarrow L^{a}$, $\Hat{\mathcal{P}} \leftarrow \mathcal{P}^{c}$\;
} 
\If{$\mathcal{U}^{a} > L_{G}$}{ 
$\mathcal{H} \leftarrow \mathcal{H} \cup (\mathcal{P}^{a}, \Bar{\mathcal{P}}, \mathcal{Y},\mathcal{U}^{a})$\;
}
$(\mathcal{P}^{c}, L^{b}, \mathcal{Y},\mathcal{U}^{b}) \leftarrow \text{BoundEstimation}(\mathcal{P}^{b}, \Bar{\mathcal{P}}, a_{z}, \mathcal{Y})$\;
\If{$L^{b} > L_{G}$}{
$L_{G} \leftarrow L^{b}$, $\Hat{\mathcal{P}} \leftarrow \mathcal{P}^{c}$\;
} 
\If{$\mathcal{U}^{b} > L_{G}$}{ 
$\mathcal{H} \leftarrow \mathcal{H} \cup (\mathcal{P}^{b}, \Bar{\mathcal{P}}, \mathcal{Y},\mathcal{U}^{b})$\;
}
}}}
return $\Hat{\mathcal{P}}$\;
  \caption{Branch and Bound Approach for Influential Billboard Slot Selection Problem}
 \label{Algo:BB}
\end{algorithm}
\vspace{-0.3in}
\paragraph{\textbf{Complexity Analysis.}}
Now, we analyze the time and space requirements for Algorithm \ref{Algo:BB}. Initialization at Line No. $1,3$ will take $\mathcal{O}(1)$ time and initialize max heap in Line No. $2$ will take $\mathcal{O}(m \cdot \log \cdot m)$ time. In-Line No. $4$ \texttt{while loop} will execute for $\mathcal{O}(n)$ till $L^{G} < U^{G}$. To pick an element in Line No. $7$ will take $\mathcal{O}(1)$, and Line No. $8$ to $11$ will take $\mathcal{O}(m)$ time to execute. In-Line No. $12$ for each slot, it will take $\mathcal{O}(\ell \cdot m^{2} \cdot t)$ time to execute. Next, in Line No. $13$ to $15$ will take $\mathcal{O}(1)$ time and Line No. $16$ to $18$ will take $\mathcal{O}(m \cdot \log \cdot m)$ time. So, Line No. $12$ to $18$ will take $\mathcal{O}(\ell \cdot m^{2} \cdot t + m \cdot \log \cdot m)$ time to execute. Similarly, Line No. $19$ to $25$ will take $\mathcal{O}(\ell \cdot m^{2} \cdot t + m \cdot \log \cdot m)$ time in the worst case. Hence, In Algorithm \ref{Algo:BB} will take total $\mathcal{O}(\ell \cdot m^{2} \cdot t + m \cdot \log \cdot m)$ time to execute the \texttt{while loop} each time. The additional space requirements for Algorithm \ref{Algo:BB} will be of $\mathcal{O}(m + \ell)$.

\subsection{ Fast Bound Estimation.}
The Fast Bound Estimation approach is introduced in Algorithm \ref{Algo:FastBoundEstimation} to generate and explore the branches in each step. A key challenge is to find a tight upper bound due to the overlap influence of the billboard slots. So, to achieve this goal, we first allocate billboard slots $\mathcal{P}^{c} = \{b_{1}, b_{2}, \ldots b_{k}\}$, where $k \subseteq \mathcal{BS}$ based on the highest influence value till the budget constraint is satisfied. Let's assume $b^{extra}$ is the next billboard slot with the highest influence value. Now, if we include $b^{extra}$ into $\mathcal{P}^{c}$ total budget is exceeded, but if not included, we will lose the cost $\mathcal{B} - \mathcal{C}(\mathcal{P}^{c})$. Therefore we add $b^{extra}$ to get the upper bound, and it is calculated as $\mathcal{U}^{a} \leftarrow \mathcal{I}(\mathcal{P}^{c}) + [\mathcal{B} - \mathcal{C}(\mathcal{P}^{c})]. \frac{\mathcal{I}(\mathcal{P}^{c} \cup ~b^{extra}) - \mathcal{I}(\mathcal{P}^{c})}{\mathcal{C}(b^{extra})}$. This Algorithm starts by initializing $p^{*}$ and zone-specific influence demand $\mathcal{Y}$. In-Line No. $2$ \texttt{for loop} will execute for all the zones an advertiser has, and in Line, No. $3$ the \texttt{while loop} will run until the budget is exhausted and zonal influence demands are satisfied. If the budget and remaining influence demand are greater than zero, then Line No. $14$ to $21$ will be executed to fulfill the demand. In-Line No. $23$ to $24$, the lower and upper bound is calculated, and in Line No. $25$ $\mathcal{P}^{c}$, $\mathcal{Y}$, $\mathcal{U}^{a}$ and $L^{a}$ is returned.

\SetKwComment{Comment}{/* }{ */}
\begin{algorithm}[ht!]
\scriptsize
\SetAlgoLined
\KwData{$\mathcal{D}$, $\mathbb{BS}$, $I()$, $z_{j}$, $\mathcal{B}$, $\mathcal{P}^{a}$, $\Bar{\mathcal{P}}$, $\mathcal{Y}$.}
\KwResult{($\mathcal{P}^{c}, L^{a}, \mathcal{Y},\mathcal{U}^{a}$)}
\begin{multicols}{2}
$p^{*} \leftarrow \emptyset$, $\mathcal{P}^{c} \leftarrow \emptyset$\;
\For {each $z_{j} \in a_{z}$}{
\While{$\mathcal{C}(\mathcal{P}^{a}) + \mathcal{C}(p^{*}) \leq \mathcal{B}$ \text{and} $\mathcal{I}(\mathcal{P}^{a} \cup \{p^{*}\}) \leq \sigma_{z_{j}}$ \text{and} $\Bar{|\mathcal{P}|} \neq \emptyset$ }{

$b \leftarrow \underset{s \in \Bar{\mathcal{P}}}{argmax} \ \mathcal{I}((\mathcal{P}^{a} \cup \{p^{*}\}) \cup \{s \})$\;

\If{ $\mathcal{C}(\mathcal{P}^{a} \cup p^{*}) + \mathcal{C}(b) \leq \mathcal{B}$}{
$p^{*} \leftarrow p^{*} \cup \{b\}$\;
$\mathcal{B} \leftarrow \mathcal{B} - \mathcal{C}(b)$\;
}
$\Bar{\mathcal{P}} \leftarrow \Bar{\mathcal{P}} \setminus \{b\}$\;
}
$\mathcal{P}^{c} \leftarrow \mathcal{P}^{c} \cup (\mathcal{P}^{a} \cup p^{*})$\;
}
\columnbreak
$p^{*} \leftarrow \emptyset$\;
\While{$\mathcal{C}(\mathcal{P}^{c}) + \mathcal{C}(p^{*}) \leq \mathcal{B}$ \text{and} $\Bar{|\mathcal{P}|} \neq \emptyset$ }{

$b \leftarrow \underset{s \in \Bar{\mathcal{P}}}{argmax} \ \mathcal{I}((\mathcal{P}^{a} \cup \{p^{*}\}) \cup \{s \})$\;

\If{ $\mathcal{C}(\mathcal{P}^{c} \cup p^{*}) + \mathcal{C}(b) \leq \mathcal{B}$}{
$p^{*} \leftarrow p^{*} \cup \{b\}$\;
$\mathcal{B} \leftarrow \mathcal{B} - \mathcal{C}(b)$\;
}
$\Bar{\mathcal{P}} \leftarrow \Bar{\mathcal{P}} \setminus \{b\}$\;
}
$\mathcal{P}^{c} \leftarrow \mathcal{P}^{c} \cup p^{*}$, $\Bar{\mathcal{P}} \leftarrow \Bar{\mathcal{P}} \setminus \mathcal{P}^{c}$\;
$b^{extra} \leftarrow \underset{s \in \Bar{\mathcal{P}}}{argmax} \ \mathcal{I}(\mathcal{P}^{c} \cup \{s \})$\;
$L^{a} \leftarrow \mathcal{I}(\mathcal{P}^{c})$, $\mathcal{U}^{a} \leftarrow \mathcal{I}(\mathcal{P}^{c}) + [\mathcal{B} - \mathcal{C}(\mathcal{P}^{c})] . \frac{\mathcal{I}(\mathcal{P}^{c} \cup ~b^{extra}) - \mathcal{I}(\mathcal{P}^{c})}{\mathcal{C}(b^{extra})}$\;
return $\mathcal{P}^{c}, L^{a}, \mathcal{Y},\mathcal{U}^{a}$
\end{multicols}
 \caption{FastBoundEstimation}
 \label{Algo:FastBoundEstimation}
\end{algorithm}
\vspace{-0.4in}

\paragraph{\textbf{Complexity Analysis.}} 
Initialization at Line No. $1$ will take $\mathcal{O}(1)$ time to execute. The \texttt{for loop} at Line No. $2$ will execute for $\mathcal{O}(\ell)$ time and $\texttt{while loop}$ at line No. $3$ will execute for $\mathcal{O}(m)$ time and in Line No. $4$, computing marginal gain will take $\mathcal{O}(\ell \cdot m^{2} \cdot t)$ time in which $t$ is the number of tuples in the trajectory database. Line No. $5$ to $11$ will execute for $\mathcal{O}(m \cdot \ell)$ time. Hence, Line No. $1$ to $12$ will take total  $\mathcal{O}(\ell \cdot m^{2} \cdot t + m \cdot \ell)$ time to execute. Similarly, Line No. $13$ to $22$ will take $\mathcal{O}(\ell \cdot m^{2} \cdot t + m \cdot \ell)$ time in the worst case. In-Line No. $23$, calculating $b^{extra}$ will take $\mathcal{O}(m \cdot t)$ and Line No. $24$ will take $\mathcal{O}(4 \cdot m \cdot t)$  time. Hence, Algorithm \ref{Algo:FastBoundEstimation} will take $\mathcal{O}(\ell \cdot m^{2} \cdot t + m \cdot \ell + m \cdot t)$ i.e., $\mathcal{O}(\ell \cdot m^{2} \cdot t)$ time in the worst case. The additional space requirements for Algorithm \ref{Algo:FastBoundEstimation} will be $\mathcal{O}(m+\ell)$.

\subsection{Bound Estimation.}
The branch and bound approach invokes Algorithm \ref{Algo:BoundEstimation} for bound estimation in each iteration. From Algorithm \ref{Algo:Greedy}, we observed that selecting billboard slots using greedy takes huge computational time to calculate the marginal gain and find the slots having the maximum gain. Although the branch and bound with Algorithm \ref{Algo:FastBoundEstimation} gives a solution based on the highest influence value in each iteration, it is not optimal. However, it reduces computational costs hugely. Motivated by this, we propose a bound estimation approach in Algorithm \ref{Algo:BoundEstimation} without traversing all the slots to estimate the bound with an approximation of $\frac{\theta}{2}(1-1/e-\epsilon)$, where $\epsilon$ is a user-defined parameter which balances the trade-off between accuracy and efficiency. Also, we set a threshold $\tau$ in Line No. $2$ by calculating maximal of $(\mathcal{I}(\mathcal{P}^{a} \cup \{ b\}) - \mathcal{I}(\mathcal{P}^{a})) / \mathcal{C}(b)$ and we progressively decrease the value of $\tau$ by a factor of $(1 + e)$ and insert an element into $p^{*}$. Here, one thing needs to be observed: Algorithm \ref{Algo:BoundEstimation} does not explore all the billboard slots to find the best solution, and when the value of $\tau$ is small enough, the algorithm terminates.

\SetKwComment{Comment}{/* }{ */}
\begin{algorithm}[ht!]
\scriptsize
\SetAlgoLined
\KwData{$\mathcal{D}$, $\mathbb{BS}$, $I()$, $z_{j}$, $\mathcal{B}$, $\mathcal{P}^{a}$, $\Bar{\mathcal{P}}$, $\mathcal{Y}$.}
\KwResult{($\mathcal{P}^{c}, L^{a}, \mathcal{Y},\mathcal{U}^{a}$)}
\begin{multicols}{2}
$p^{*} \leftarrow \emptyset$, $\mathcal{P}^{c} \leftarrow \emptyset$, $\epsilon \leftarrow 0.1$\;
$\tau \leftarrow \underset{b \in \Bar{\mathcal{P}}}{max} \ \frac{\mathcal{I}(\mathcal{P}^{a} \cup \{ b\}) - \mathcal{I}(\mathcal{P}^{a})}{\mathcal{C}(b)}$, $\mathcal{B}^{r} \leftarrow \mathcal{B} - \mathcal{C}(\mathcal{P}^{a})$\;
\For {each $z_{j} \in a_{z}$}{
\While{$\mathcal{C}(\mathcal{P}^{a}) + \mathcal{C}(p^{*}) \leq \mathcal{B}$ \text{and} $\mathcal{I}(\mathcal{P}^{a} \cup \{p^{*}\}) \leq \sigma_{z_{j}}$ \text{and} $\Bar{|\mathcal{P}|} \neq \emptyset$ }{
\For{$ b \in \Bar{\mathcal{P}}$}{
\If{$\mathcal{C}(\mathcal{P}^{a} \cup p^{*}) + \mathcal{C}(b) \leq \mathcal{B}$}{
m = $\mathcal{I}(\mathcal{P}^{a} \cup \{ b\} | \mathcal{P}^{a}) - \mathcal{I}(p^{*} | \mathcal{P}^{a})$\;
\If{$\frac{m}{\mathcal{C}(b)} \geq \tau$}{
$\Bar{\mathcal{P}} \leftarrow \Bar{\mathcal{P}} \setminus \{b\}$\;
$p^{*} \leftarrow p^{*} \cup \{b\}$\;
}
\If{$\frac{m}{\mathcal{C}(b)} < \tau$}{
break\;
}}}
$\tau \leftarrow \frac{\tau}{1+\epsilon}$\;
\If{$\tau \leq \frac{\mathcal{I}(p^{*} | \mathcal{P}^{a})}{\mathcal{B}^{r}} \cdot \frac{e^{-1}}{1-e^{-1}}$}{
break\;
}}
$\mathcal{P}^{c} \leftarrow \mathcal{P}^{c} \cup (\mathcal{P}^{a} \cup p^{*})$\;
}
\columnbreak
$p^{*} \leftarrow \emptyset$, $\Bar{\mathcal{P}} \leftarrow \Bar{\mathcal{P}} \setminus \mathcal{P}^{c}$, $\mathcal{B} \leftarrow \mathcal{B} - \mathcal{C}(\mathcal{P}^{c})$\;

\While{$\mathcal{C}(\mathcal{P}^{c}) + \mathcal{C}(p^{*}) \leq \mathcal{B}$ \text{and} $\Bar{|\mathcal{P}|} \neq \emptyset$ }{
\For{$ b \in \Bar{\mathcal{P}}$}{
\If{$\mathcal{C}(\mathcal{P}^{c} \cup p^{*}) + \mathcal{C}(b) \leq \mathcal{B}$}{
m = $\mathcal{I}(\mathcal{P}^{c} \cup \{ b\} | \mathcal{P}^{c}) - \mathcal{I}(p^{*} | \mathcal{P}^{c})$\;
\If{$\frac{m}{\mathcal{C}(b)} \geq \tau$}{
$\Bar{\mathcal{P}} \leftarrow \Bar{\mathcal{P}} \setminus \{b\}$\;
$p^{*} \leftarrow p^{*} \cup \{b\}$\;
}
\If{$\frac{m}{\mathcal{C}(b)} < \tau$}{
break\;
}}}
$\tau \leftarrow \frac{\tau}{1+\epsilon}$\;
\If{$\tau \leq \frac{\mathcal{I}(p^{*} | \mathcal{P}^{a})}{\mathcal{B}^{r}} \cdot \frac{e^{-1}}{1-e^{-1}}$}{
break\;
}}
$\mathcal{P}^{c} \leftarrow \mathcal{P}^{c} \cup p^{*}$, $\Bar{\mathcal{P}} \leftarrow \Bar{\mathcal{P}} \setminus \mathcal{P}^{c}$\;
$b^{extra} \leftarrow \underset{s \in \Bar{\mathcal{P}}}{argmax} \ \frac{\mathcal{I}(\mathcal{P}^{c} \cup \{s \}) - \mathcal{I}(\mathcal{P}^{c})}{\mathcal{C}(\{s\})}$\;
$L^{a} \leftarrow \mathcal{I}(\mathcal{P}^{c})$, $\mathcal{U}^{a} \leftarrow \mathcal{I}(\mathcal{P}^{c}) + [\mathcal{B} - \mathcal{C}(\mathcal{P}^{c})] . \frac{\mathcal{I}(\mathcal{P}^{c} \cup ~b^{extra}) - \mathcal{I}(\mathcal{P}^{c})}{\mathcal{C}(b^{extra})}$\;
return $\mathcal{P}^{c}, L^{a}, \mathcal{Y},\mathcal{U}^{a}$
\end{multicols}
 \caption{BoundEstimation}
 \label{Algo:BoundEstimation}
\end{algorithm}
\vspace{-0.3in}
\paragraph{\textbf{Complexity Analysis.}}
Initialization at Line No. $1$ and calculating $\tau$ at Line No. $2$ will take $\mathcal{O}(1)$ and $\mathcal{O}(2 \cdot m \cdot t)$ time, respectively. Line No. $3$ \texttt{for loop} and Line No. $4$ \texttt{while loop} will execute for $\mathcal{O}(\ell)$ and $\mathcal{O}(m)$ time, respectively. Line No. $5$ \texttt{for loop} will execute for $\mathcal{O}(m^{2} \cdot \ell)$ and Line No. $7$ will take $\mathcal{O}(2 \cdot m^{2} \cdot \ell \cdot t)$ time to execute. Line No. $8$ to $16$ will execute for $\mathcal{O}(m^{2} \cdot \ell)$ and Line No. $17$ will execute for $\mathcal{O}(m \cdot \ell)$. Finally, in Line No. $18$ to $20$ will take $\mathcal{O}(m \cdot \ell \cdot t)$ time. Hence, Line No. $1$ to $23$ will take $\mathcal{O}(2 \cdot m \cdot t + 2 \cdot m^{2} \cdot \ell \cdot t + m^{2} \cdot \ell + m \cdot \ell + m \cdot \ell \cdot t)$ i.e., $\mathcal{O}(2 \cdot m^{2} \cdot \ell \cdot t)$ time to execute. Similarly, Line No. $24$ to $43$ will take $\mathcal{O}(2 \cdot m^{2} \cdot \ell \cdot t)$ time to execute. In-Line No. $44$ will take $\mathcal{O}(m \cdot t)$ and Line No. $45$ will take $\mathcal{O}(4 \cdot m \cdot t)$ time to execute. Hence, Algorithm \ref{Algo:BoundEstimation} will take $\mathcal{O}(2 \cdot m^{2} \cdot \ell \cdot t + 2 \cdot m^{2} \cdot \ell \cdot t + 2 \cdot m \cdot t + 4 \cdot m \cdot t)$ i.e., $\mathcal{O}(m^{2} \cdot \ell \cdot t)$ time in the worst case to execute. The additional space requirements for Algorithm \ref{Algo:BoundEstimation} will be $\mathcal{O}(m+\ell)$.

\paragraph{\textbf{Theoretical analysis related to performance.}}
Due to space limitations, we have not conducted an elaborate analysis of our proposed methods. In our case, we have used the results from the study by Zhang et al. \cite{10.1145/3292500.3330829} and state their result in Theorem \ref{Th: Analysis}.

\begin{figure}[!h]
    \begin{minipage}{0.4\textwidth}
        \centering
        \begin{tabular}{| c | c | c | c | c |}
        \hline
        $\mathcal{BS}_{i}$ & $bs_{1}$ & $bs_{2}$ & $bs_{3}$ & $bs_{4}$ \\ \hline
        $I(bs_{i})$ & 2 & 3 & 7 & 5  \\ \hline
        $C(bs_{i})$ & 100 & 200 & 400 & 300 \\ \hline
        Zone Info. & $z_{1}$ & $z_{1}$ & $z_{2}$ & $z_{3}$  \\ \hline
        \end{tabular}
        \caption{\label{ETable:1} Billboard Info.}
    \end{minipage}%
    \begin{minipage}{0.45\textwidth}
        \centering
        \includegraphics[width=70mm]{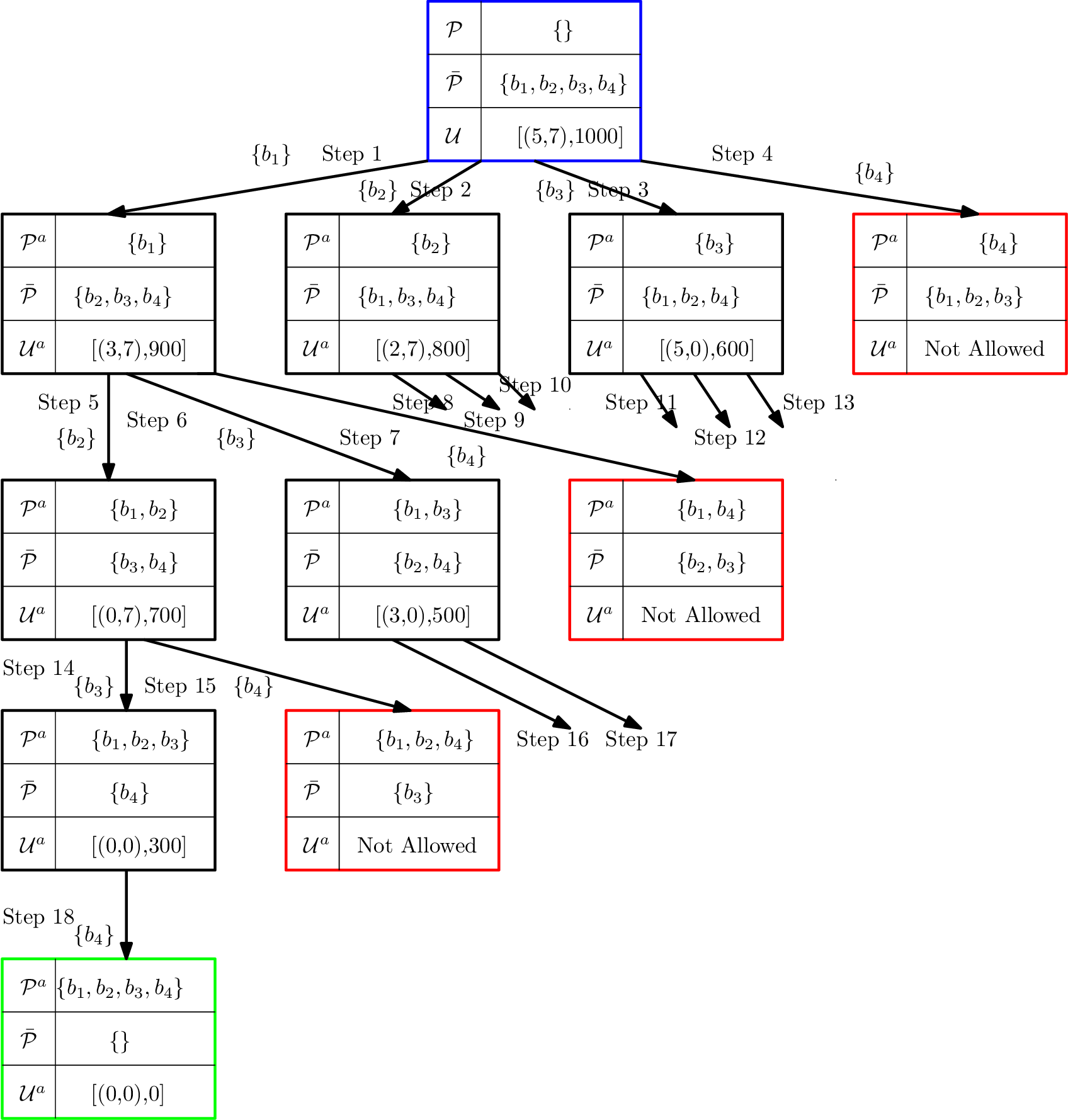}
        \caption{A running example}
        \label{Fig:1}
    \end{minipage}
\end{figure}

\begin{theorem}\label{Th: Analysis}
    The branch and bound approach invoking Algorithm \ref{Algo:BoundEstimation} and it results an optimal solution set $(\mathcal{OPT})$ from the billboard slots set $\Bar{\mathcal{P}}$ then it follows the following criteria : 
    \begin{equation}
         \mathcal{I}(\Bar{\mathcal{P}}) \geq \frac{\theta}{2}(1-1/e-\epsilon) \cdot \mathcal{I}(\mathcal{OPT}).
    \end{equation}
\end{theorem}
\paragraph{\textbf{An Illustrative Example.}}
Consider an influence provider has four billboard slots $\mathcal{BS} = \{ b_{1}, b_{2}, b_{3}, b_{4}\}$ and an advertiser $a_{1}$ approaches to the influence provider with required influence demand from two different zones $z_{1}$ and $z_{2}$ for budget $\mathcal{B}$. It can be denoted in the form of vector $[(\sigma_{z_{1}},\sigma_{z_{2}}), \mathcal{B}]$. Now, among these four slots $b_{1}$ and $b_{2}$ belongs to zone $z_{1}$ with influence value $2$ and $3$ respectively, $b_{3}$ belongs to zone $z_{2}$ with influence value $7$ and $b_{4}$ belongs to zone $z_{3}$ with influence value $5$ as shown in Table \ref{ETable:1}. Next, the working of Branch and Bound framework in Algorithm \ref{Algo:BB} is as follows: At first, we initialize the max heap $\mathcal{H}$ with an entry $(\{\}, \{b_{1},b_{2},b_{3},b_{4}\}, [(5,7),1000])$. The set $\mathcal{P}$ is empty, and the set $\Bar{\mathcal{P}}$ contains all the billboard slots $\{ b_{1}, b_{2}, b_{3}, b_{4}\}$ in the top part of the search tree as shown in the Figure \ref{Fig:1}. In the search tree, every node has an entry like $(\mathcal{P}, \Bar{\mathcal{P}}, \mathcal{U})$ and in each step the values of $\mathcal{P}$, $\Bar{\mathcal{P}}$, and $\mathcal{U}$ are updated. Next, in step $1$, we pop the initial heap entry and, based on the budget and zonal constraint, push a new branch into the heap, i.e., $(\{b_{1}\}, \{b_{2},b_{3},b_{4}\}, [(3,7),900])$ by invoking Algorithm \ref{Algo:FastBoundEstimation} or \ref{Algo:BoundEstimation}. Since the $L^{a} > L_{G}$, we update the value of $\Hat{\mathcal{P}}$ with $\mathcal{P}^{c}$. Similarly, at steps $2,3,4$, the newly generated branches are pushed into the heap. However, in step $4,7,10,13,15,17$, adding billboard slot $b_{4}$ does not satisfy the zonal influence constraint, and these entries are not inserted into the heap. Next, in step $14$, the zonal influence demand is satisfied, but some budget remains. So, this budget is utilized by taking slots from available zones, i.e., $b_{4}$ is added to the solution set, and the budget is exhausted. From Figure \ref{Fig:1}, it is clear that the path ($step~1 \rightarrow ~ step ~ 5 \rightarrow step ~14 \rightarrow step ~18$) gives an optimal solution that satisfies all the constraints and when $L_{G} \geq U_{G}$ Algorithm \ref{Algo:BB} terminates.
 \vspace{-0.1 in}
\vspace{-0.1in}

\section{Experimental Evaluation} \label{Sec:EE}
This section discusses the experimental evaluation to show the effectiveness of the proposed solution methodologies. Initially, we start by describing the datasets.
\paragraph{\textbf{Dataset Descriptions.}}
We conduct experiments on real-world trajectory check-in datasets for New York City (NYC) and Los Angeles (LA). The NYC\footnote{\url{https://www.nyc.gov/site/tlc/about/tlc-trip-record-data.page}} dataset contains $227,428$ check-ins between April 12, 2012, and February 16, 2013, and The LA\footnote{\url{https://github.com/Ibtihal-Alablani}} dataset contains $74,170$ check-in from $15$ different streets of Los Angles. We have crawled billboard datasets from the different locations of New York City and Los Angeles from LAMAR\footnote{\url{http://www..lamar.com/InventoryBrowser}}, one of the largest billboard providers worldwide. The NYC billboard dataset consists of $1031040$ slots and the LA dataset consists of $2135520$  slots. However, in our experiment, we used $1500$ and $2500$ slots for NYC and LA, respectively. Additionally, we divide the billboard datasets based on latitude and longitude for New York City and Los Angeles into $5$ and $3$ different geographic zones, respectively.

\subsection{Experimental Setup.}
The following experimental setups are considered in our experiments.
\paragraph{\textbf{Key Parameters.}}
All the key parameters are summarized in Table \ref{Table-2}, and the default parameters are highlighted in bold. In our experiments, we vary only one parameter and keep the rest of the parameters in the default setting. In Algorithm \ref{Algo:BB}, $\theta$ controls the termination condition while $\epsilon$ is used to make trade efficiency with accuracy. The distance threshold $\eta$ denotes the maximum distance in which a billboard can influence the maximum number of trajectories.
\begin{table}[h!]
\caption{\label{Table-2} Key Parameters}
\vspace{-0.15 in}
\begin{center}
    \begin{tabular}{ | p{2cm}| p{5.5cm}|}
    \hline
    Parameter & Values  \\ \hline
    $\mathcal{B}$ & $5k, 10k, 15k, 20k, 25k$   \\ \hline
    $\epsilon$ & $10^{-4}, 10^{-3}, 10^{-2}, 10^{-1}, 1$ \\ \hline
    $\theta$ & $0.5, 0.6, \textbf{0.7}, 0.8, 0.9$  \\ \hline
    $\eta$ & $25m,50m,\textbf{100m},125m,150m$  \\ \hline
    \end{tabular}
\end{center}
\end{table}
\paragraph{\textbf{Environment Setup.}}
All Python codes were executed on an HP Z4 workstation equipped with 64 GB of memory and an Xeon(R) 3.50 GHz processor.
\paragraph{\textbf{Billboard cost.}}
An advertising company like LAMAR does not provide an exact cost of billboard slots. As reported in the existing studies \cite{zahradka2021price,10.1145/3219819.3219946,10.1145/3292500.3330829}, a billboard's cost is proportional to its corresponding influence value. Hence, we have calculated using the following setting: $\mathcal{C}(b) = \lfloor\delta * \mathcal{I}(b) / 10 \rfloor$, where $\delta$ is a factor randomly chosen between $0.8$ to $1.1$ to simulate the cost fluctuation.
\paragraph{\textbf{Performance Measurement.}}
Each baseline and the proposed method are evaluated for their influence and runtime, and each experiment is repeated five times to demonstrate the effectiveness and efficiency of the proposed methods. Average results from these experiments are reported.
\subsection{Algorithms Compared.} In this paper, our goal is to evaluate the effectiveness and efficiency of the proposed methods. Hence, we compare our methods with the following baselines:
\paragraph{\textbf{Top-$k$.}}
Each iteration selects a billboard slot, which influences a maximum number of trajectories until the zone-specific influence demand and budget constraints are satisfied.
\paragraph{\textbf{Random}.} This is the simple approach where it will choose billboard slots uniformly at random in each iteration until the zonal influence and budget constraints are satisfied.
\subsection{Goals of the Experiments.}
In this study, we want to address the following Research Questions (RQ).
\begin{itemize}
\item \textbf{RQ1}: How do the influence value and runtime change if we increase the budget to select billboard slots?
\item \textbf{RQ2}: For the Branch and Bound algorithm, varying values of $\theta$ and $\epsilon$, how do the runtime and the influence quality change?
\item \textbf{RQ3}: How do the influence value and run-time change if the number of demanding zones increases by the advertisers?
\item \textbf{RQ4}: How do the influence and run-time change if we vary trajectory size?
\end{itemize}

\subsection{\textbf{Experimental Results.}}
This section will discuss experimental results for the proposed and existing solution methodologies. Due to space limitations, we cannot include different varying parameter results, which will come in the journal version of this work. 

\begin{figure*}[!ht]
\centering
\begin{tabular}{ccc}
\includegraphics[scale=0.17]{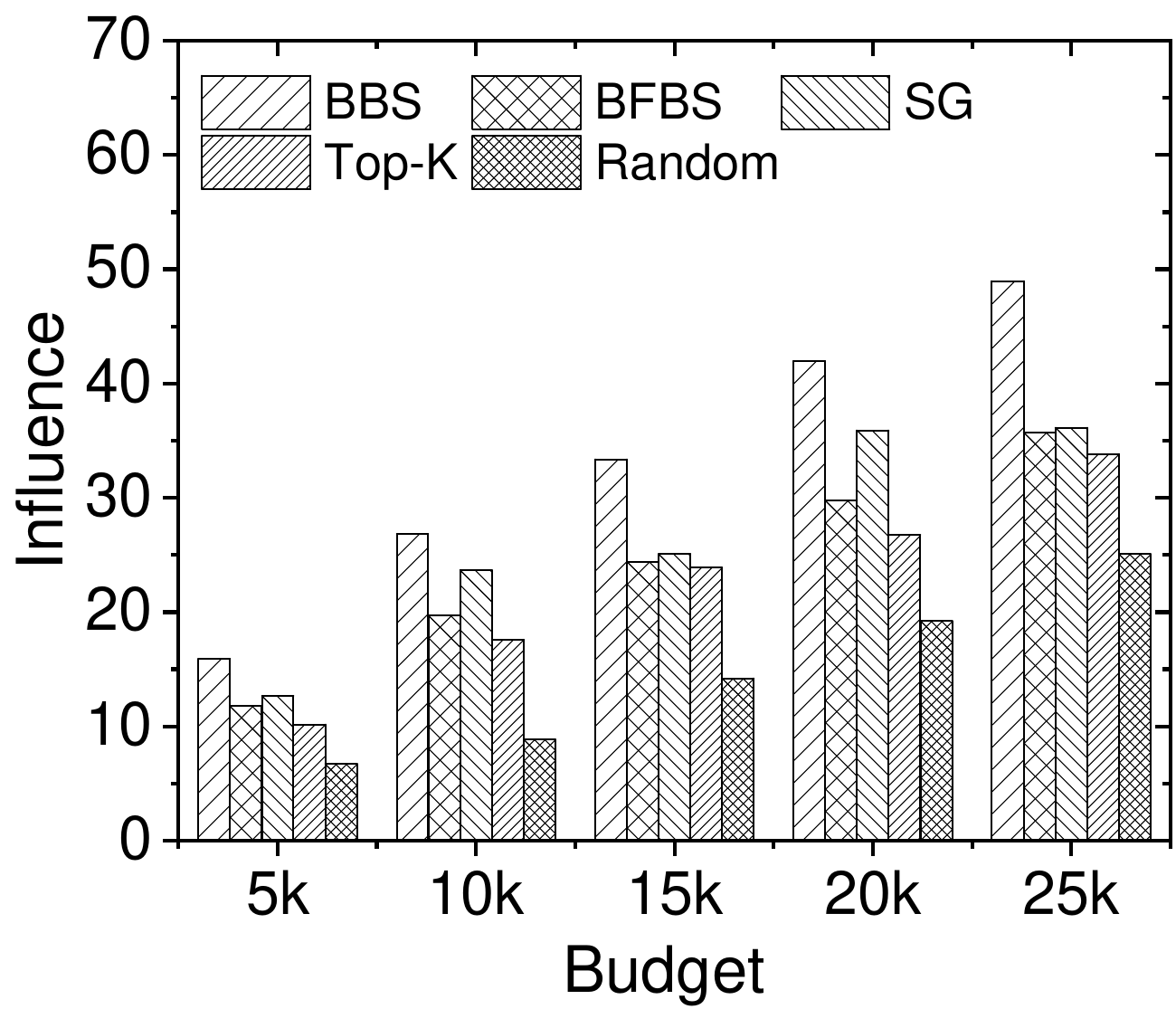} & \includegraphics[scale=0.17]{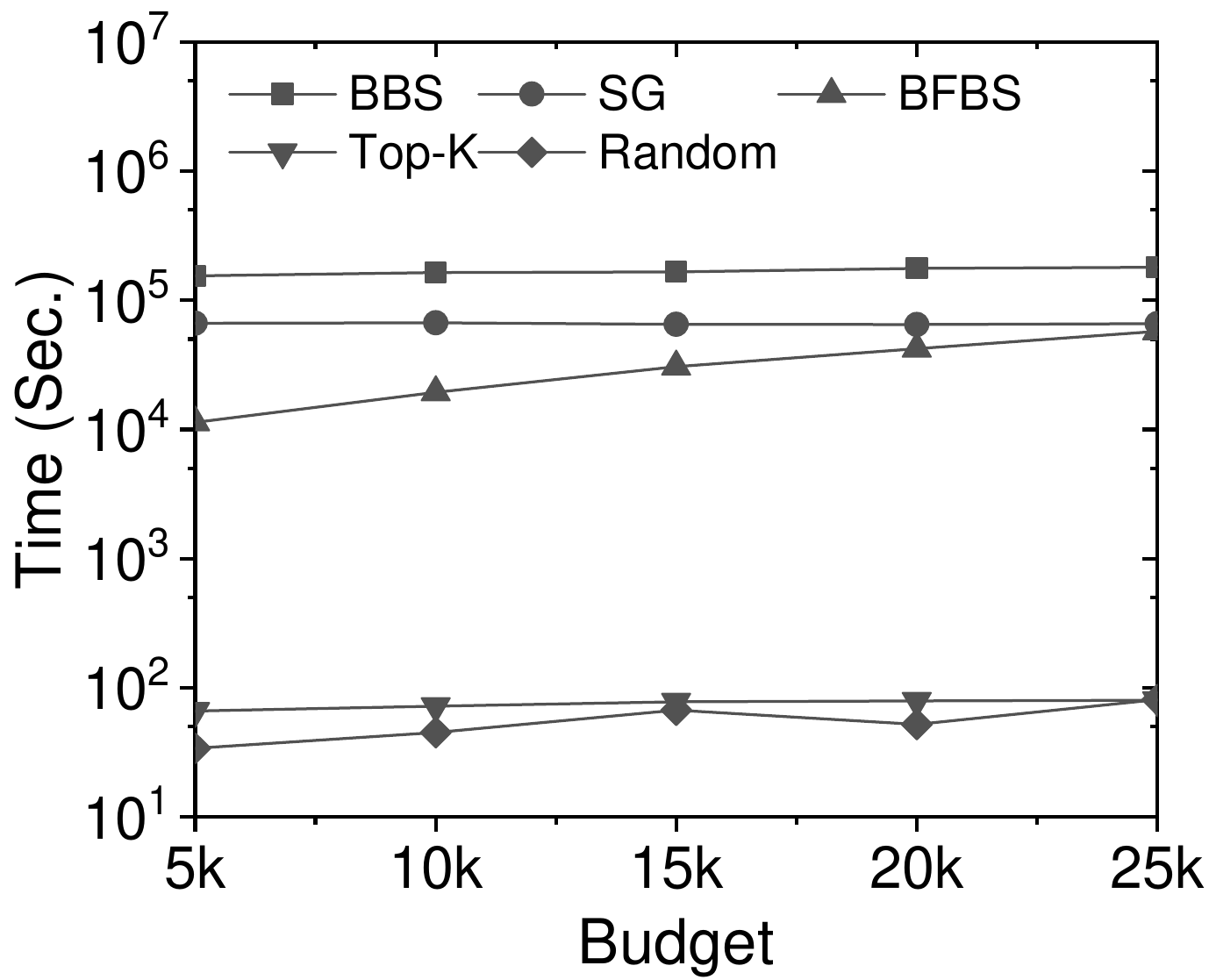} & \includegraphics[scale=0.17]{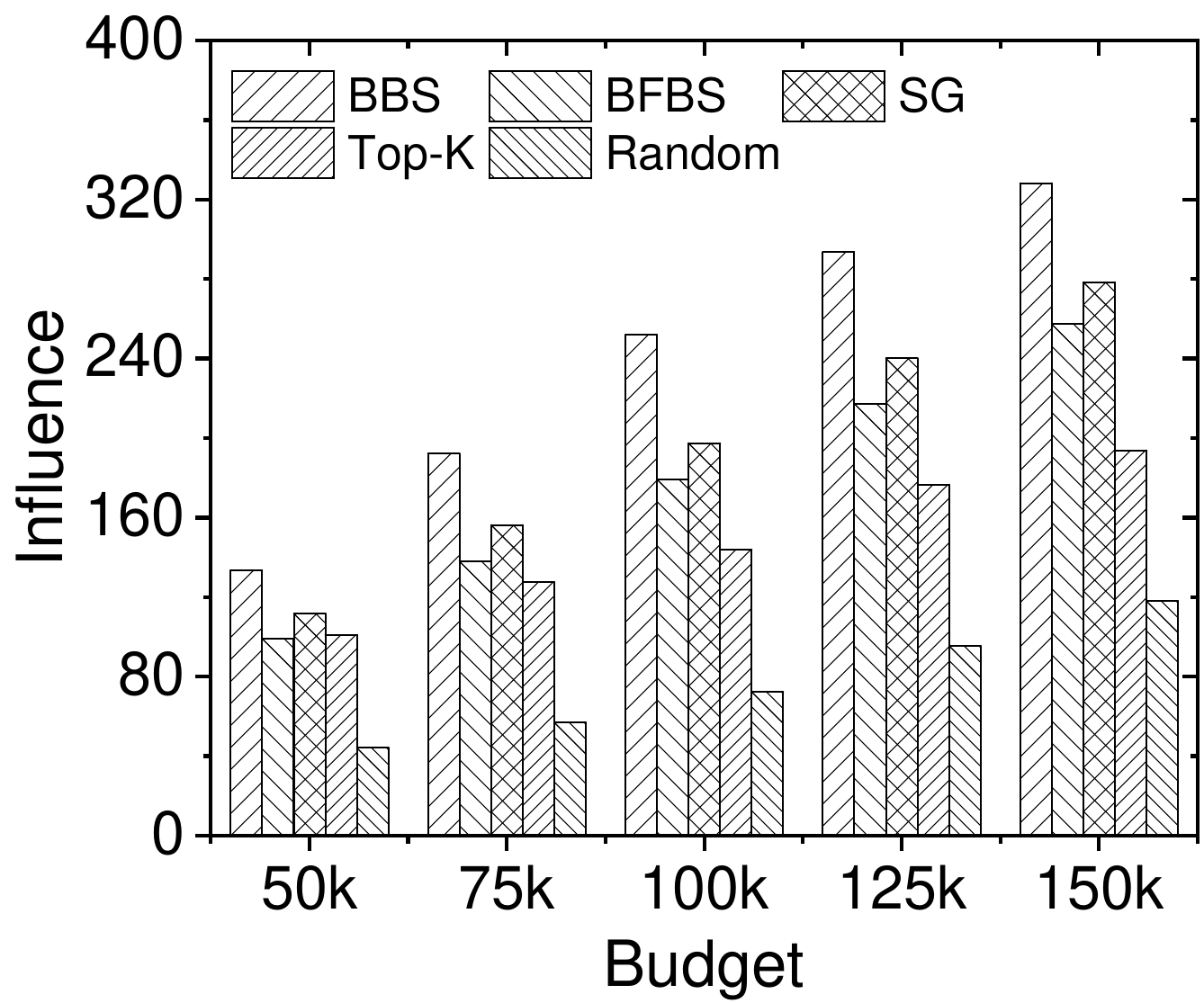} \\
(a) Influence in NYC & (b) Runtime in NYC & (c) Influence in LA  \\
\includegraphics[scale=0.17]{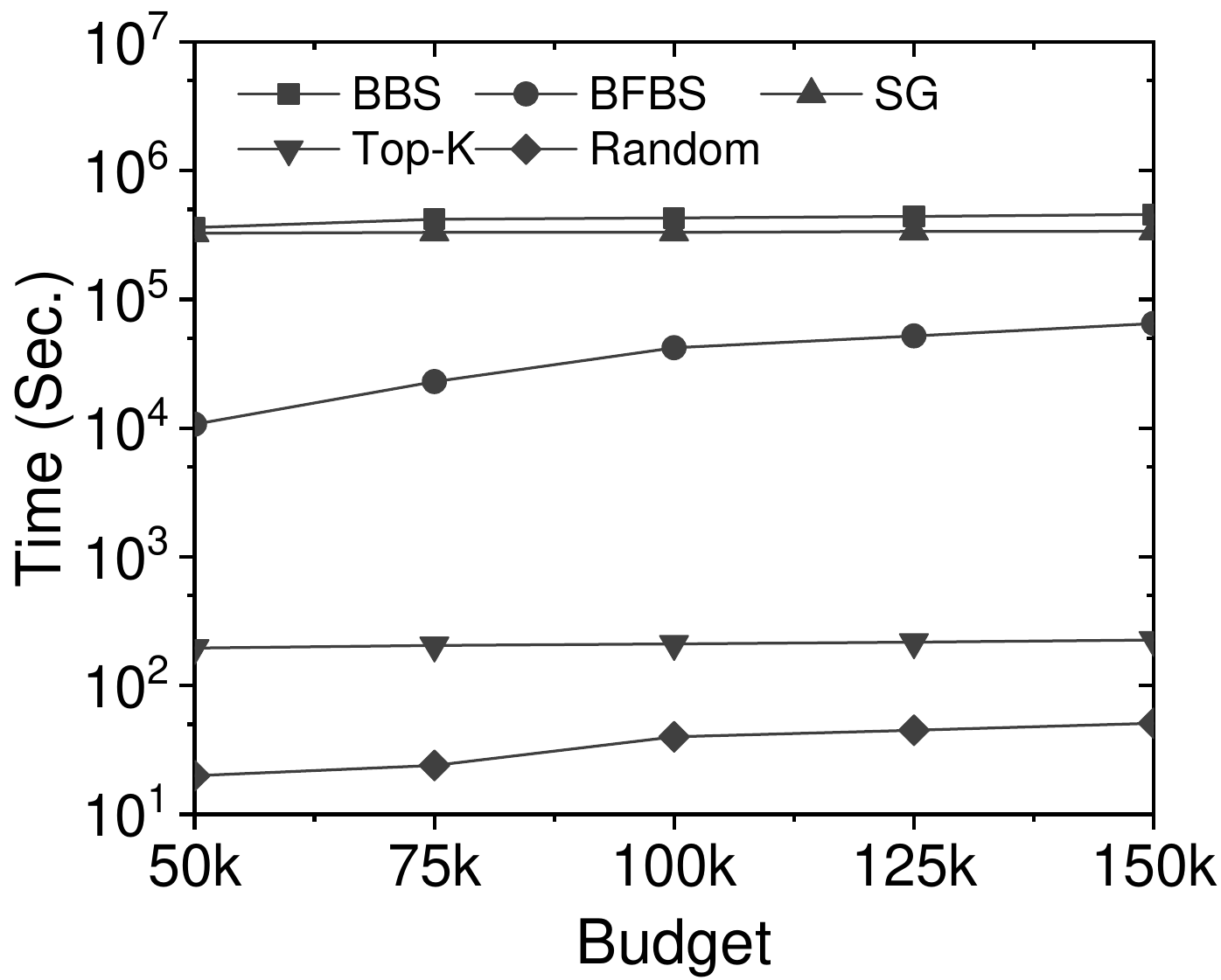}  & \includegraphics[scale=0.17]{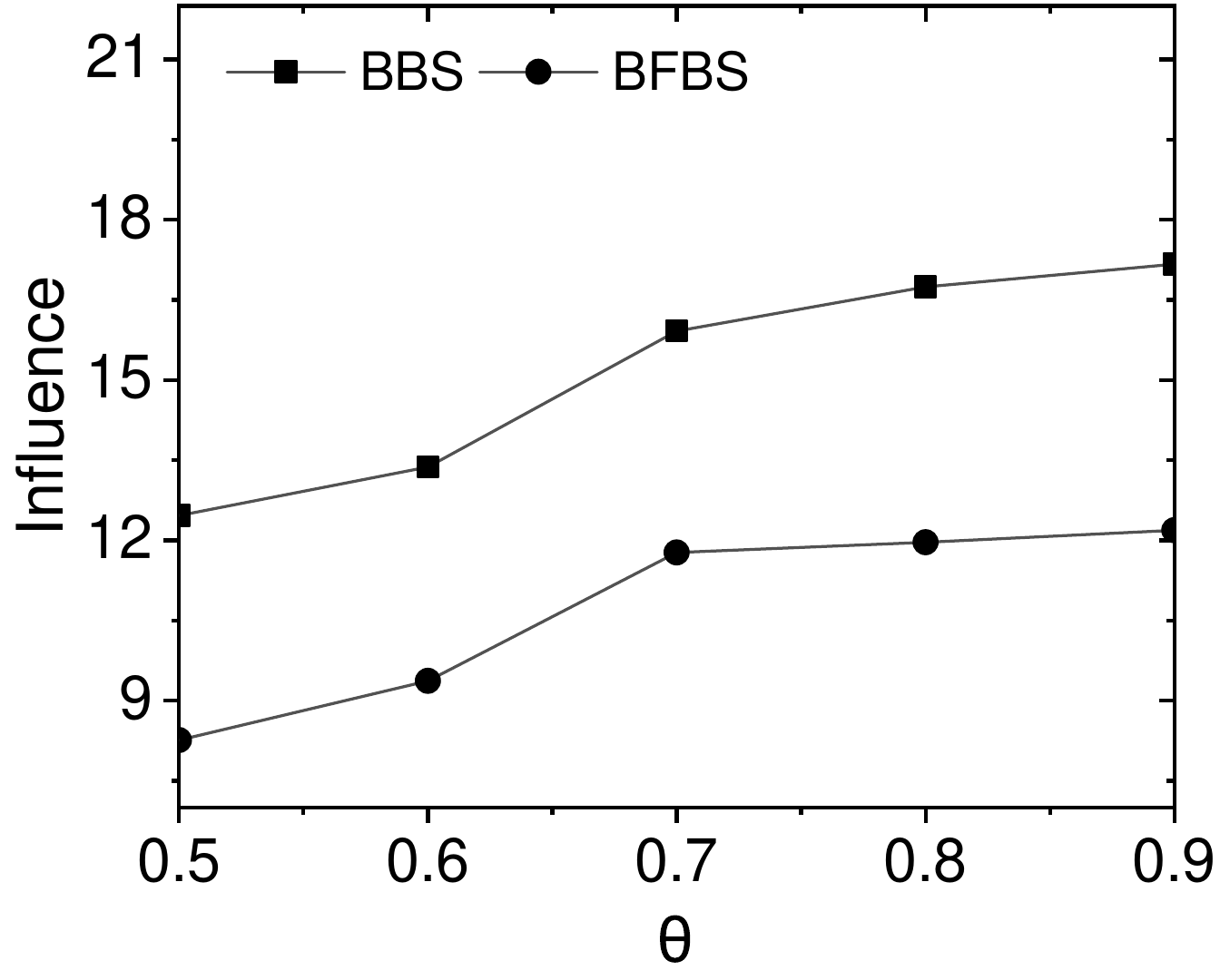} & \includegraphics[scale=0.17]{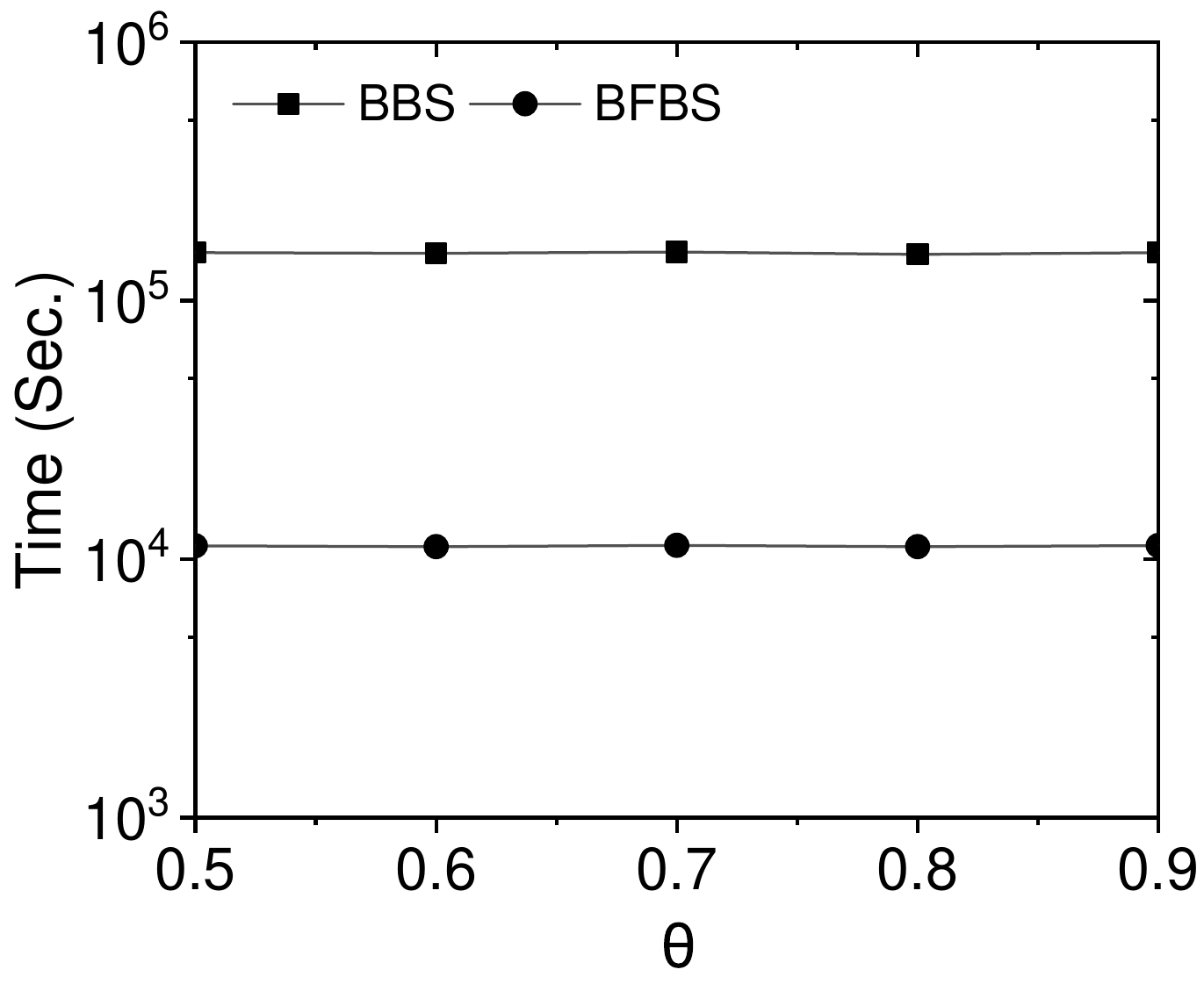}  \\
(d) Runtime in LA & (e) $\theta$ Vs. Influence (NYC) & (f) $\theta$ Vs. Time (NYC) \\
\includegraphics[scale=0.17]{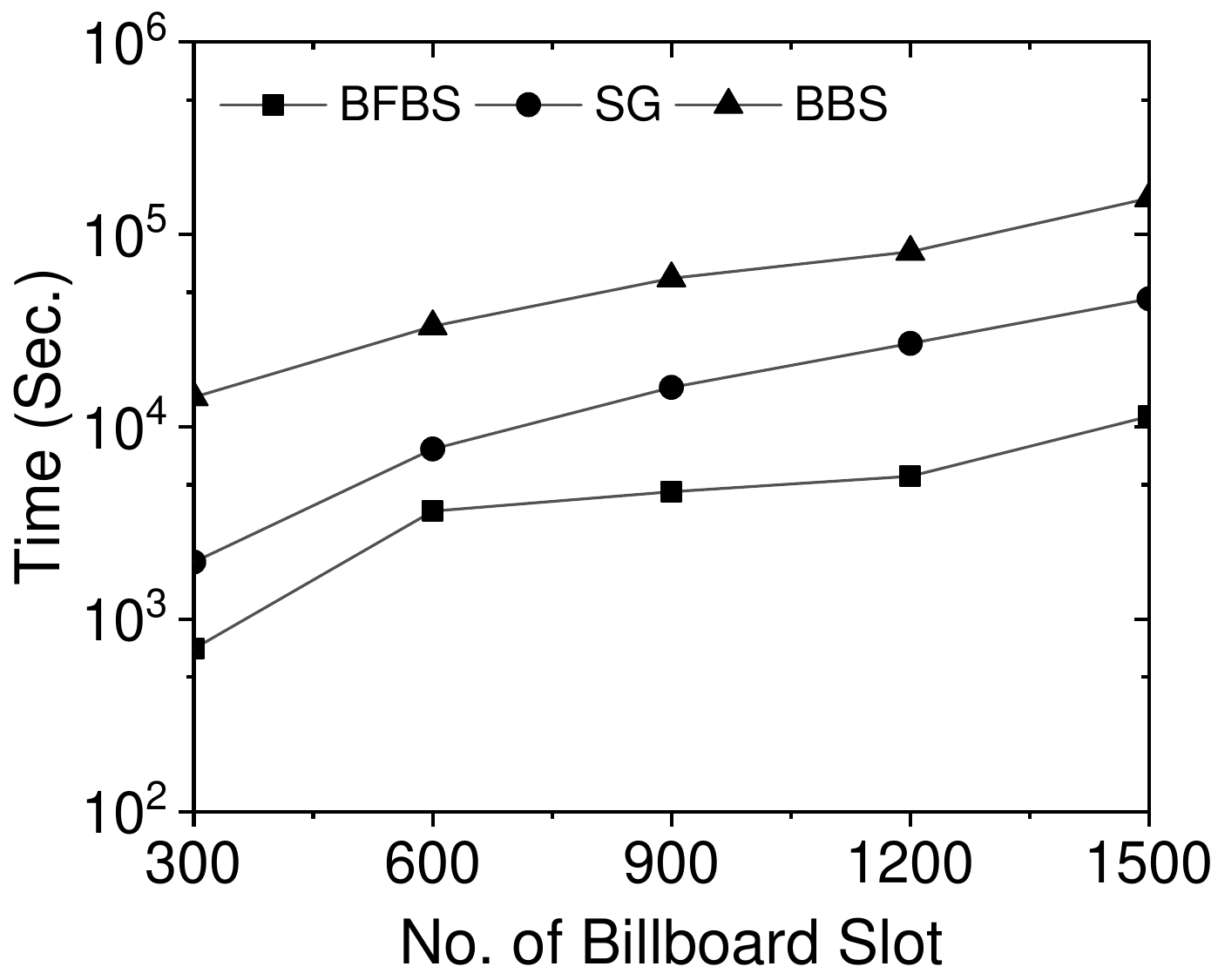} & \includegraphics[scale=0.17]{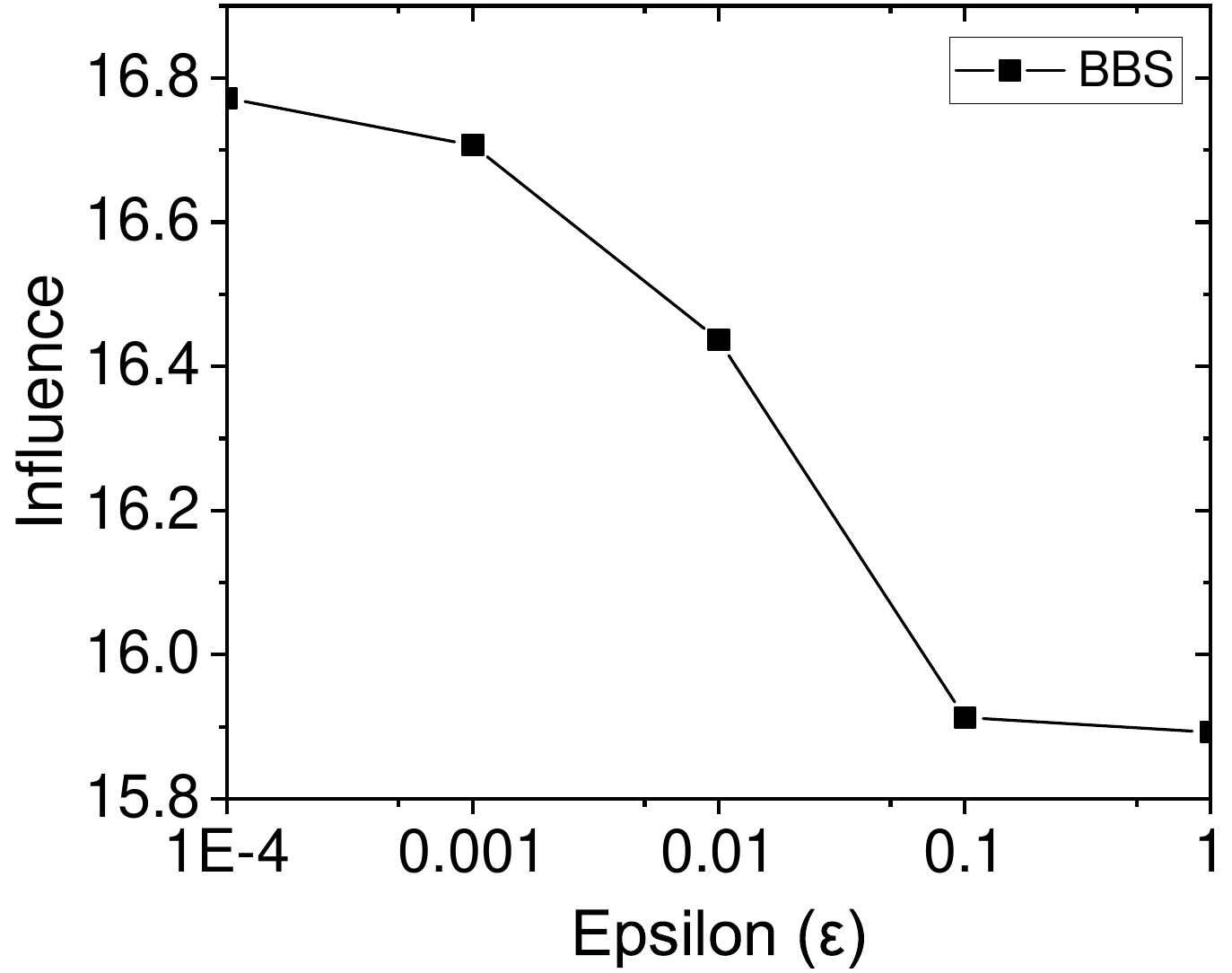} & \includegraphics[scale=0.17]{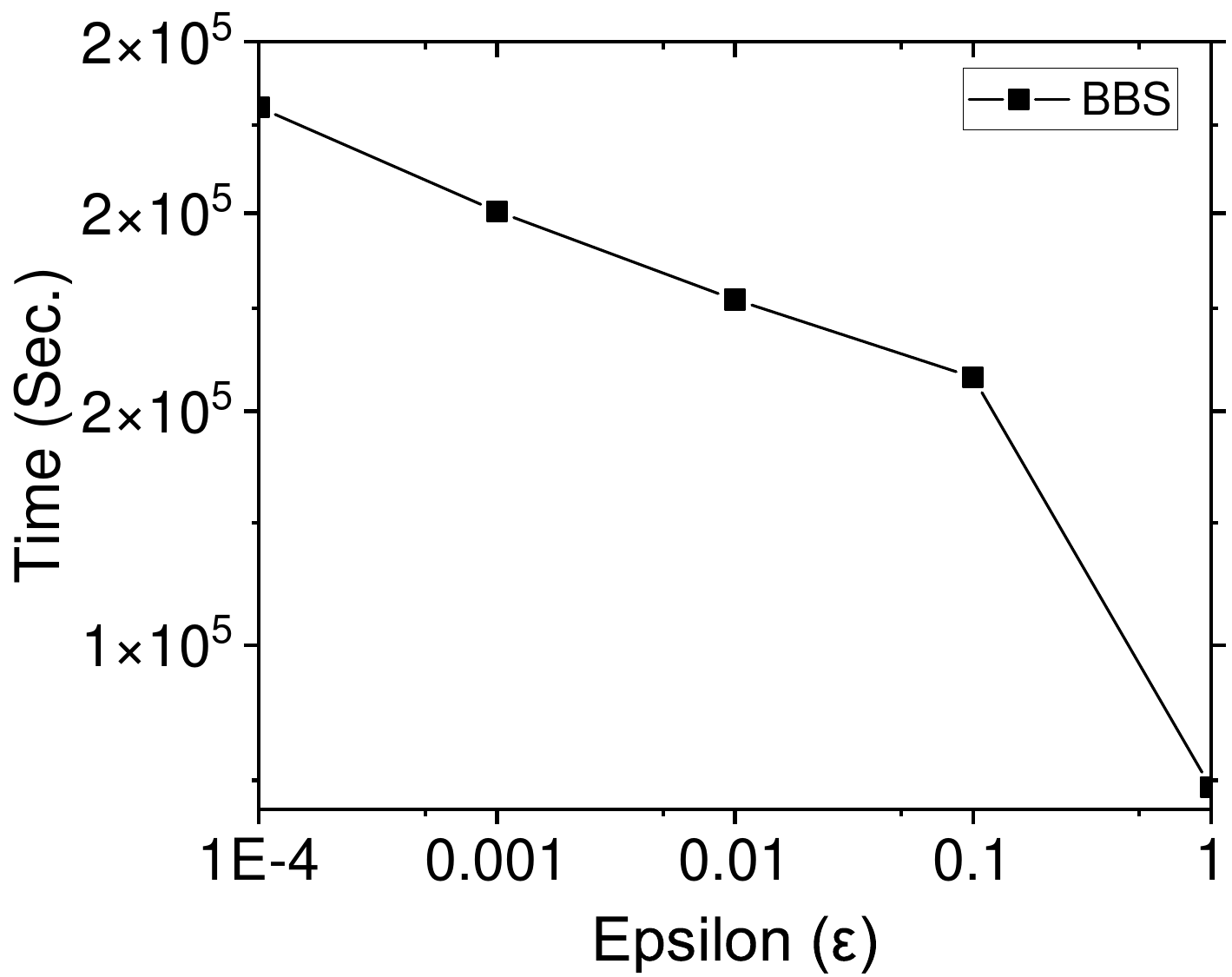}  \\
(g) Scalability Test (NYC) & (h) $\epsilon$ Vs. Influence (NYC) & (i) $\epsilon$ Vs. Time (NYC) \\
\end{tabular}
\caption{Varying Budget in NYC (a,b), LA (c,d), Varying $\theta$ in NYC (e,f), Varying No. of Slot in NYC (g), Varying $\epsilon$ in NYC (h,i)}
\label{Fig:Influence}
\end{figure*}

\paragraph{\textbf{Budget Vs. Influence.}}
Figure \ref{Fig:Influence}(a) and \ref{Fig:Influence}(c) show the effectiveness of the proposed as well as baseline methodologies in the NYC and LA datasets. From this, we have significant observations. First, when the budget increases from $5k$ to $25k$ in NYC and $50k$ to $150k$ in the LA dataset, the BBS outperforms the BFBS and the greedy approach by $26\%$ to $28\%$ and $20\%$ to $27\%$ respectively.
Second, the BFBS has worse performance, as it gives preference to the slots with the highest influence value in each iteration, which are the most expensive in the real world. So, it can choose only a limited number of slots when the budget is less, and due to the influence overlap, the BBS performs well compared to the BFBS approach. Third, the impact of the BBS and the BFBS is greater in NYC than in the LA dataset, and the possible reason is the distribution of trajectory in the trajectory dataset, i.e., the influence overlap is less in the NYC Dataset. 
\paragraph{\textbf{Budget Vs. Time.}}
From Figure \ref{Fig:Influence}(b) and \ref{Fig:Influence}(d), we have three observations. First, the running time of BFBS is very low compared to the BBS and Greedy approach because in BFBS, for every branch, the slots are selected based on the highest individual influence value, whereas in Greedy, slots are selected based on marginal gain computation. However, for every branch, the BBS method compute marginal gain for each slot in each iteration, which resulted in a significant increase in run time. Second, with the increased budget, the computational cost of each proposed and baseline method increases. Third, Top-$k$ and Random is the fastest approach because they scan all billboard slots only once.

\paragraph{\textbf{Varying $\epsilon, \theta$.}}
The user-defined parameter $\theta$ controls the termination criteria in Algorithm \ref{Algo:BB}. The larger $\theta$ value will increase the effectiveness with worse efficiency. Figure \ref{Fig:Influence}(c), \ref{Fig:Influence}(d) shows the effect of varying $\theta$ in the NYC dataset. Next, when the $\epsilon$ value increases from $0.0001$ to $1$, the effectiveness of the BFBS and BBS decreases by at least $5\%$ to $10\%$.However, the efficiency increases by $20\%$ to $30\%$. We found that the effectiveness and efficiency are stable when $\theta$ and $\epsilon$ values are $0.7$ and $0.1$, respectively, and chose this as the default setting. 

\paragraph{\textbf{Scalability Test.}}
Figure \ref{Fig:Influence}(g) shows that in the NYC dataset, efficiency is very sensitive in BBS compared to the BFBS when we vary the billboard slot size from $300$ to $1500$. It is observed that the run-time increase rate in BBS compared to Greedy and BFBS is almost $57\%$ to $86\%$ and $92\%$ to $95\%$, respectively. 
\paragraph{\textbf{Additional Discussion.}}
Now, we discuss the impact of additional parameters, e.g., varying distance $(\eta)$, no. of zones $(|\ell|)$, size of the trajectory $(|\mathcal{T}|)$, etc. First, the influence of all the proposed and baseline methods increases with the distance increase. This happens because each billboard slot can influence a large number of trajectories. Second, when zone-specific influence demand rises with the fixed budget, the computational cost of BBS and BFBS decreases. This happens because most of the budget is used to select slots for demanding zones, and very little is used to choose slots from the remaining slots in any of the zones. Third, in the case of the NYC dataset varying trajectory size from $40k$ to $120k$, the computational time increases drastically in BBS compared to the BFBS and Greedy approach. The BBS and Greedy outperform the BFBS and baseline methods by up to $40\%$ to $60\%$ in terms of influence with the increase in trajectory size.

\section{Conclusion and Future Research Direction} \label{Sec:Conclusion}
In this paper, we study the problem of influential billboard slot selection under zonal influence constraint and formulate this as a discrete optimization problem. We propose a simple greedy and branch and bound approach with two bound estimation techniques to address this problem. We experimented with two real-world billboards and trajectory datasets and analyzed their time and space requirements. We also analyzed proposed methods to obtain the performance guarantee. Our proposed BBS gives an optimal solution; however, its computational time requirements are its bottleneck. Therefore, our future work on this problem is to develop more efficient solution methodologies.


%
%
%
\bibliographystyle{splncs04}
\bibliography{Paper}

\end{document}